\documentclass[3p,times,procedia]{elsarticle}
\usepackage{nupha_ecrc}

\volume{00}

\firstpage{1}

\journalname{Nuclear Physics A}

\runauth{R. Belmont}

\jid{nupha}

\jnltitlelogo{Nuclear Physics A}

\usepackage{amssymb}

\usepackage[figuresright]{rotating}

\newcommand{\dndeta}{dN_{ch}/d\eta}

\newcommand{\la}{\langle}
\newcommand{\ra}{\rangle}

\newcommand{\mean}[1]{\la #1 \ra}
\newcommand{\dmean}[1]{\la\la #1 \ra\ra}

\newcommand{\vtt}{v_2\{2\}}
\newcommand{\vtf}{v_2\{4\}}
\newcommand{\vtg}{v_2\{2,|\Delta\eta|>2\}}
\newcommand{\ctt}{c_2\{2\}}
\newcommand{\ctf}{c_2\{4\}}

\newcommand{\snn}{\sqrt{s_{_{NN}}}}

\newcommand{\nfvtxt}{N_{\rm tracks}^{\rm FVTX}}

\begin{document}

\begin{frontmatter}




\title{PHENIX results on multiparticle correlations in small systems}


\author{R.~Belmont (for the PHENIX Collaboration)}

\address{University of Colorado Boulder, Boulder, CO 80309-0390, USA}

\begin{abstract}
We present measurements of 2- and 4-particle correlations in d+Au collisions at four
different center-of-mass energies: 200, 62.4, 39, and 19.6 GeV. The data were collected in
2016 by the PHENIX experiment at RHIC. The second Fourier coefficient $v_2$ of the particle
azimuthal distributions is measured using the Q-cumulant method as a function of event
multiplicity. The results give a strong indication of collective behavior down to the
lowest energy.
\end{abstract}


\end{frontmatter}


\section{Introduction}
\label{sec:intro}

The study of collective behavior in small systems, like p+Pb at the LHC and d+Au at RHIC,
is one of the major pillars of heavy-ion physics research.
Measurements of collectivity in small systems are testing the limits of collectivity
and the applicability of hydrodynamics.
In 2016, RHIC delivered a beam
energy scan of d+Au collisions at four different collisions energies: 200 GeV, 62.4 GeV,
39 GeV, 19.6 GeV.
%
Table~\ref{tab:events} shows the number of events analyzed for each collision energy.
We used a central trigger that greatly enhanced the high multiplicity
data sample.

\begin{table}[h!]
\begin{center}
\begin{tabular}{rrr}
\hline
d+Au collision energy & total events analyzed & central events analyzed\\
\hline
200 GeV & 636 million & 585 million \\
62.4 GeV & 131 million & 76 million \\
39 GeV & 137 million & 49 million \\
19.6 GeV & 15 million & 3 million \\
\hline
\end{tabular}
\end{center}
\caption{Number of events analyzed by PHENIX}
\label{tab:events}
\end{table}

Measurements of multiparticle correlations in small systems at the LHC
(see e.g. Reference~\cite{Aad:2013fja,Chatrchyan:2013nka,Abelev:2014mda})
are considered very strong evidence for collective behavior, as by mathematical construction
they reduce contributions from few particle correlations like resonance decays, quantum
correlations, Coulomb interactions, momentum conservation effects, etc, generally referred
to as non-flow.  In this proceedings, we present PHENIX results on multiparticle
correlations in the 2016 d+Au beam energy scan.  We will also present results from simulations
using A Multi-Phase Transport model (AMPT)~\cite{Lin:2004en} to aid interpretations
of the data as needed.

\section{Analysis}
\label{sec:analysis}

The azimuthal distribution of particles produced in a heavy-ion collision
can be described with a Fourier expansion
with coefficients $v_n$~\cite{Voloshin:1994mz}.  In this analysis we focus on the second
harmonic coefficient $v_2$.
To study multiparticle correlations, we use the Q-cumulant method~\cite{Bilandzic:2010jr}.
To briefly summarize the salient features, the 2-particle cumulant can be described as
$\ctt = \mean{v_2^2}$
and the 4-particle cumulant can be described as
$\ctf = \mean{v_2^4} - 2\mean{v_2^2}^2$,
where the brackets indicate an average over events.
The harmonic coefficients can be determined as
$\vtt = (\ctt)^{1/2}$
and
$\vtf = (-\ctf)^{1/4}$.
The individual components can be obtained from
$\mean{v_2^2} = \dmean{2}$
and
$\mean{v_2^4} = \dmean{4}$,
where $\mean{2}$ and $\mean{4}$ are 2- and 4-particle correlators
averaged over all particles in a single event
(calculated using Q-vectors),
which are then averaged over events to
obtain $\dmean{2}$ and $\dmean{4}$.

Since neither $\vtt$ nor $\vtf$ measure $\mean{v_2}$ directly, they are
subject to fluctuations in the $v_2$ distribution.  Generally we have
$\vtt = (\mean{v_2}^2+\sigma^2)^{1/2}$
and
$\vtf \approx (\mean{v_2}^2-\sigma^2)^{1/2}$~\cite{Ollitrault:2009ie}, where $\sigma^2$ is the
variance of the $v_2$ distribution.  We also expect fewer particle correlations to be more
subject to non-flow, so that
$\vtt = (\mean{v_2}^2+\sigma^2+\delta^2)^{1/2}$~\cite{Ollitrault:2009ie}, where $\delta^2$
parameterizes the non-flow contribution.  In this way, the study of correlations with different
numbers of particles can potentially elucidate the relationship between $v_2$, the
fluctuations, and the non-flow effects.

We use the PHENIX forward vertex (FVTX) detector for this analysis.
The FVTX has a nominal pseudorapidity coverage of $1<|\eta|<3$.
Reconstructed tracks are required to have hits and at least 3 of the 4 layers
and to have a distance of closest approach to the vertex of
$|{\rm DCA}|<$~2~cm.
Events with a collision vertex of $|z_{\rm vertex}|<$~10~cm
are selected.
We plot all quantities as a function of the number of reconstructed
tracks, $\nfvtxt$.

\section{Results}
\label{sec:results}

\begin{figure}[h!]
\includegraphics[width=0.99\linewidth]{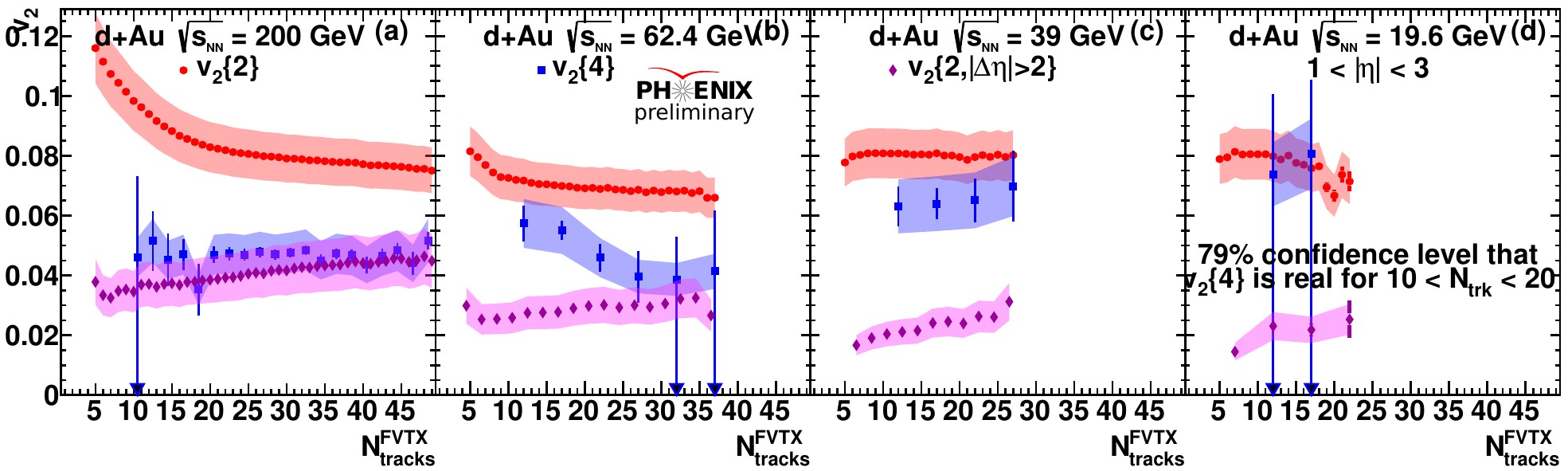}
\caption{$\vtt$, $\vtg$, and $\vtf$ as a function of $\nfvtxt$ in d+Au collisions at
  200 GeV (a), 62.4 GeV (b), 39 GeV (c), 19.6 GeV (d).}
\label{fig:4panel}
\end{figure}

Figure~\ref{fig:4panel} shows $\vtt$ in red points and $\vtf$ in blue points as a function
of $\nfvtxt$ in d+Au collisions at 200 GeV (panel a), 62.4 GeV (panel b), 39 GeV (panel
c), and 19.6 GeV (panel d).  We observe real valued $\vtf$ at all four
energies (albeit with 79\% confidence interval for the 19.6 GeV).
Additionally, we observe that although $\vtt > \vtf$ for all energies, the
difference decreases as the collision energy decreases.  This result may be
regarded as surprising,
because naively it would appear to indicate that the variance of the distribution
is decreasing significantly with decreasing collision energy
(we note that there are other possibilities, beyond the scope of these proceedings).

\begin{figure}[h!]
\begin{center}
\includegraphics[width=0.5\linewidth]{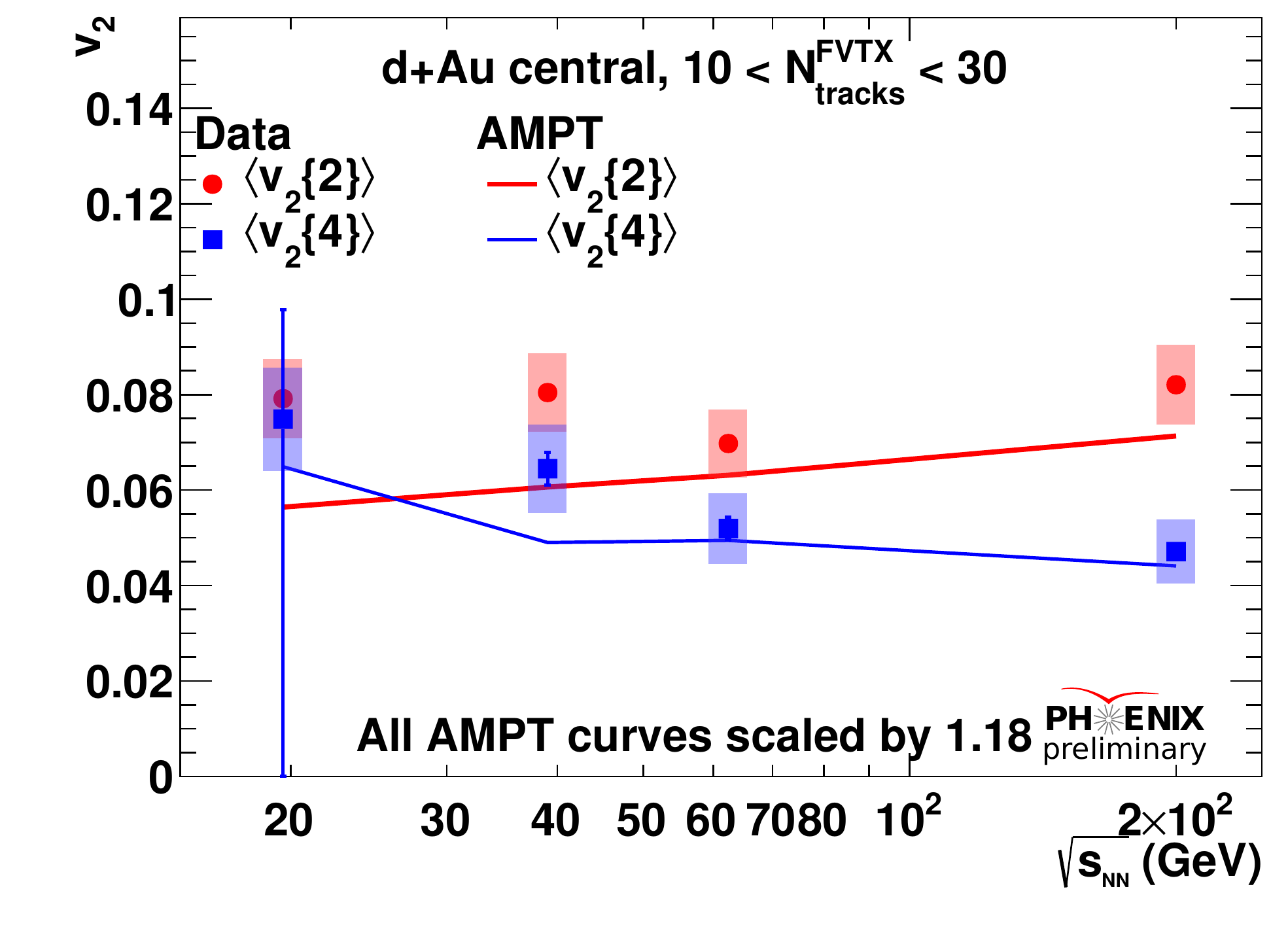}
\end{center}
\caption{$\vtt$ and $\vtf$ as a function of $\snn$ in d+Au collisions.  The data are indicated
  as points and the AMPT simulations are shown as lines.}
\label{fig:avgsqrts}
\end{figure}

Because AMPT has had significant success in describing collective observables in both large and small systems,
we can check if AMPT shows a similar trend.  Figure~\ref{fig:avgsqrts} shows
$\vtt$ and $\vtf$ as a functions of collision energy $\snn$.  The data are shown in points
and the AMPT results are shown as lines.
The analysis of the AMPT simulations is done in exactly the same way as for the real data,
including and especially the pseudorapidity selection,
to ensure a homogeneous comparison.
It can be seen that AMPT reproduces this trend.
The AMPT simulations are scaled by a factor of 1.18 to account for the $p_T$ dependence of the
FVTX efficiency.

\begin{figure}[h!]
\includegraphics[width=0.50\linewidth]{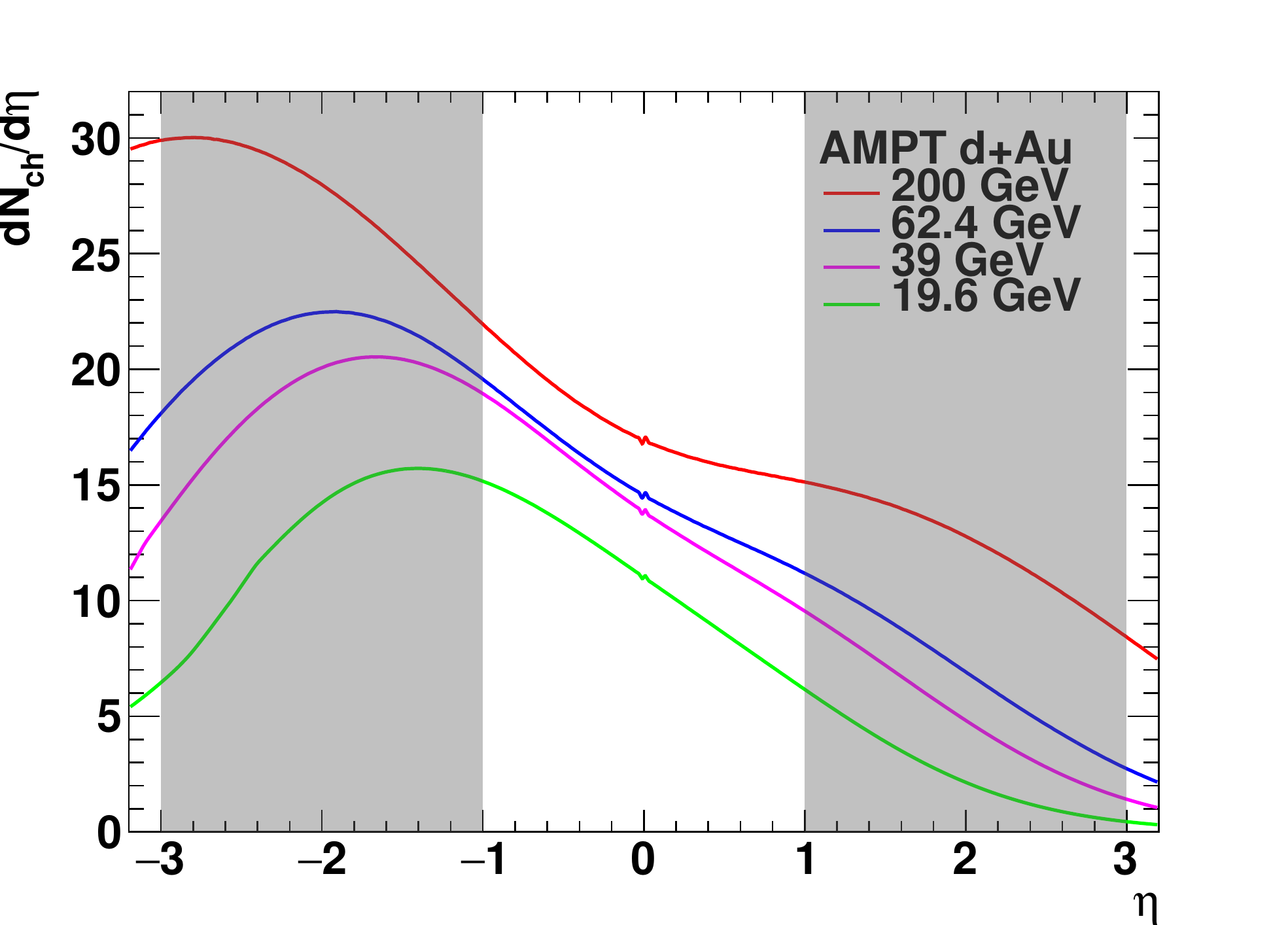}
\includegraphics[width=0.50\linewidth]{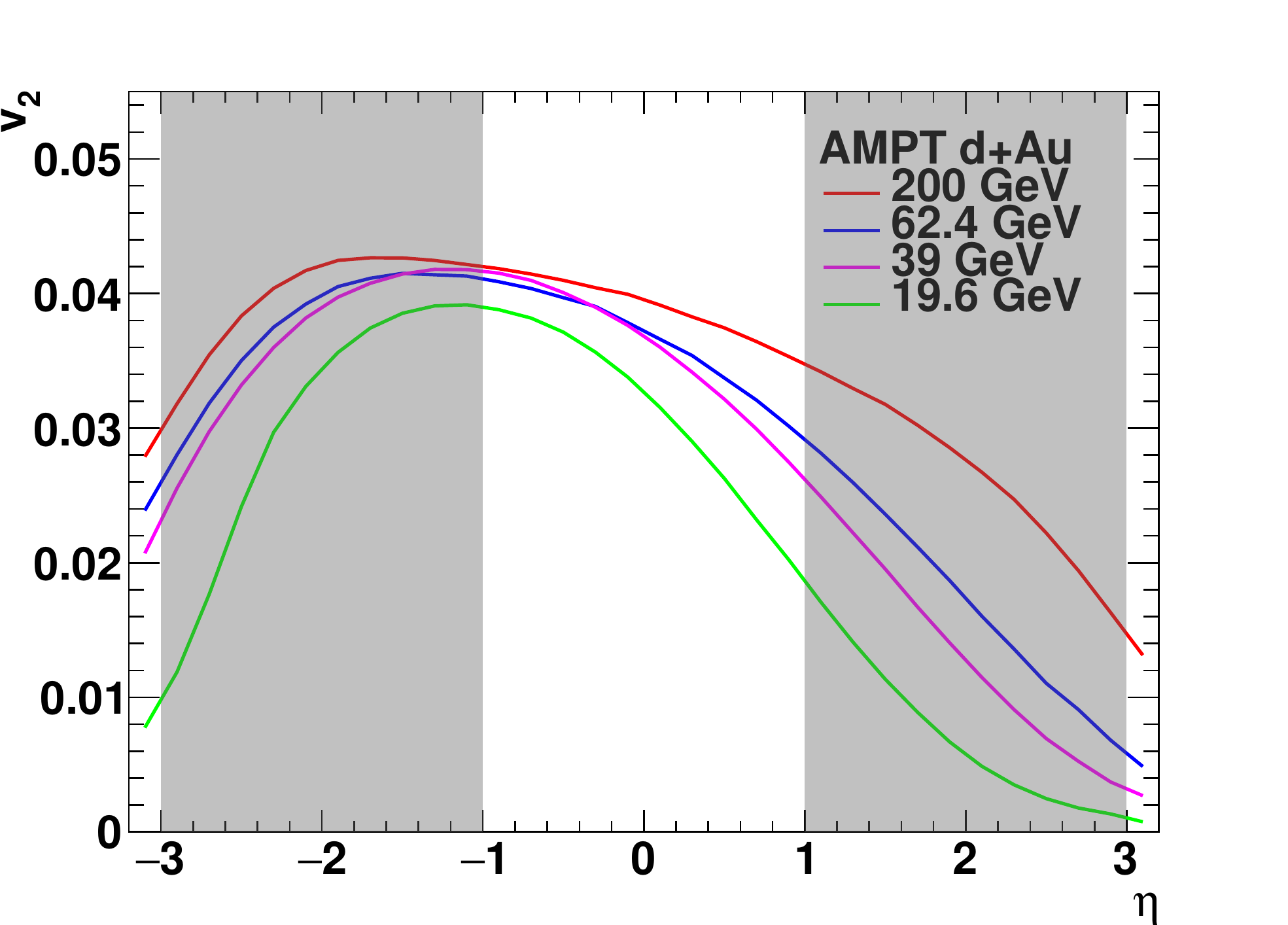}
\caption{AMPT simulations of $\dndeta$ (left) and $v_2$ vs $\eta$ in d+Au
  collisions at 200, 62.4, 39, and 19.6 GeV.  The shaded bands indicated the
  nominal FVTX acceptance.}
\label{fig:ampteta}
\end{figure}

To try to assess how the interplay of fluctuations and non-flow effects influence
the relationship between $\vtt$ and $\vtf$, we can also look at 2-particle $v_2$
with a pseudorapidity gap to reduce the non-flow.
Figure~\ref{fig:4panel} also shows $\vtg$, where $|\Delta\eta|>2$
indicates that the minimum pseudorapidity separation (gap) is 2 units.
To achieve this pseudorapidity gap,
we require one particle to be in the south (backward pseudorapidity)
arm of the FVTX and the other to be in the north (forward pseudorapidity) arm.
We find that $\vtg < \vtf$ for all four energies, with the difference increasing
as the collision energy decreases.  This can be understood as arising from the fact
that $\vtt$ and $\vtf$ are weighted averages of the backward pseudorapidity $v_2^B$
and the forward pseudorapidity $v_2^F$ whereas the $\vtg$ gives equal weight to each,
$\vtg = \sqrt{v_2^Bv_2^F}$.  In asymmetric collisions,
$\dndeta$~\cite{Alver:2010ck}
and
$v_2$~\cite{v2eta}
are
larger at backward pseudorapidity.  As an illustrative example, Figure~\ref{fig:ampteta}
shows AMPT simulations of $\dndeta$ (left panel) and $v_2$ (right panel) as a
function of $\eta$.  Based on this alone, we can plausible expect $\vtg < \vtf$.
However, it is likely that longitudinal decorrelations~\cite{Xiao:2012uw}
play a role in further reducing $\vtg$, where we have something like
$\vtg = \sqrt{v_2^Bv_2^F\cos(2(\psi_2^B-\psi_2^F))}$,
where $\psi_2^B$ and $\psi_2^F$ indicate the 2\textsuperscript{nd} harmonic event planes at
backward and forward pseudorapidity, respectively.

\begin{figure}[h!]
\includegraphics[width=0.33\linewidth]{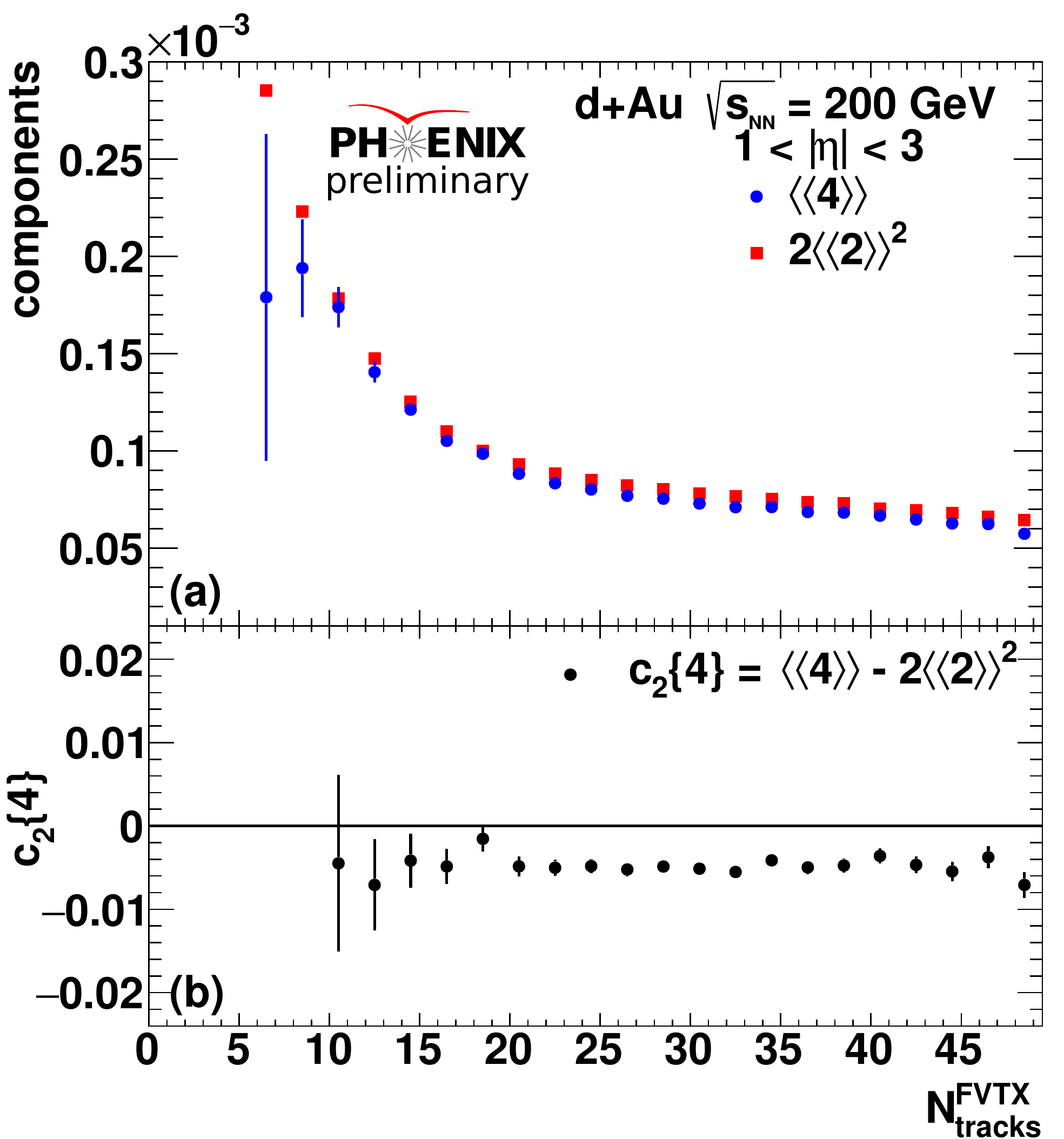}
\includegraphics[width=0.33\linewidth]{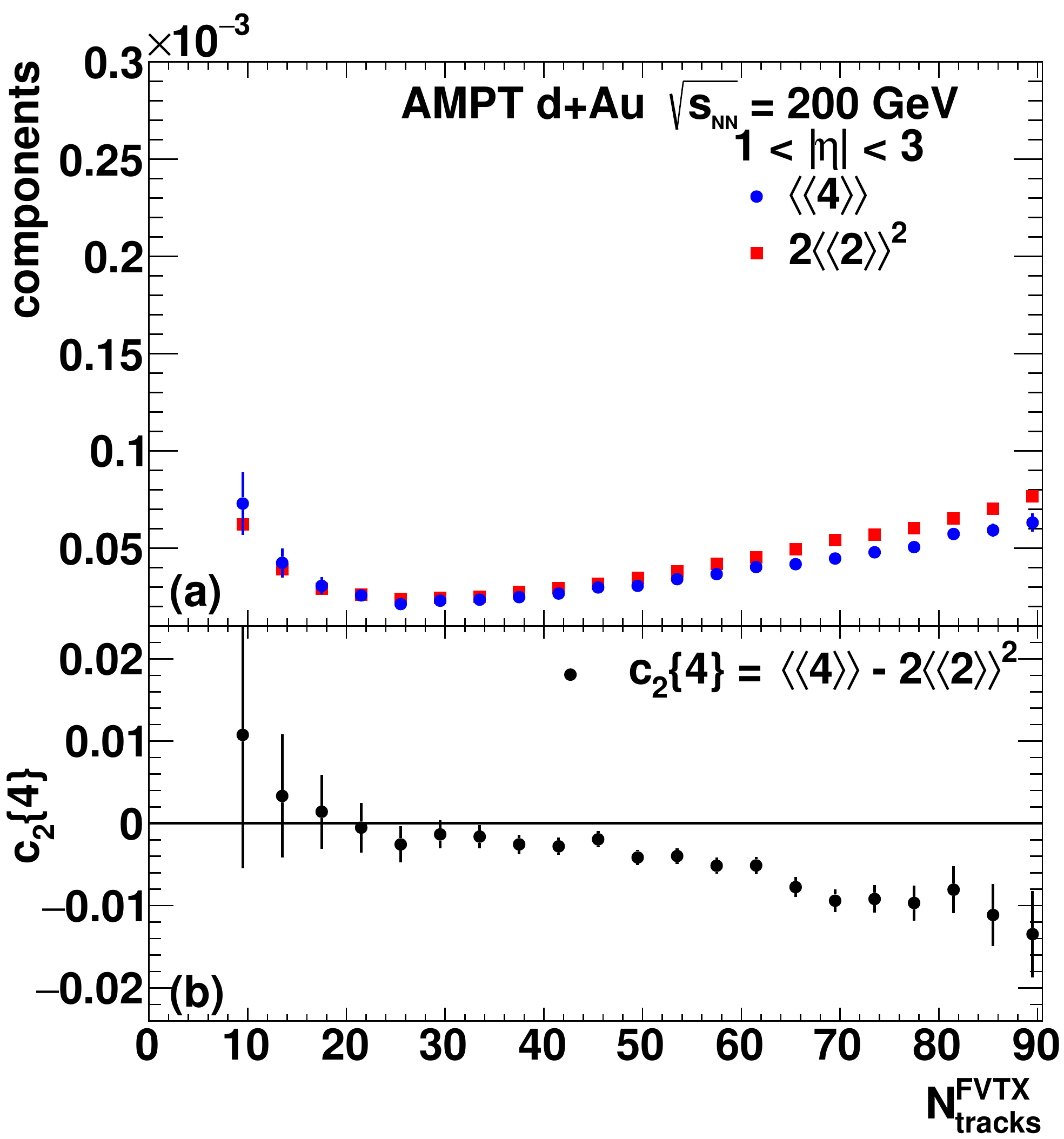}
\includegraphics[width=0.33\linewidth]{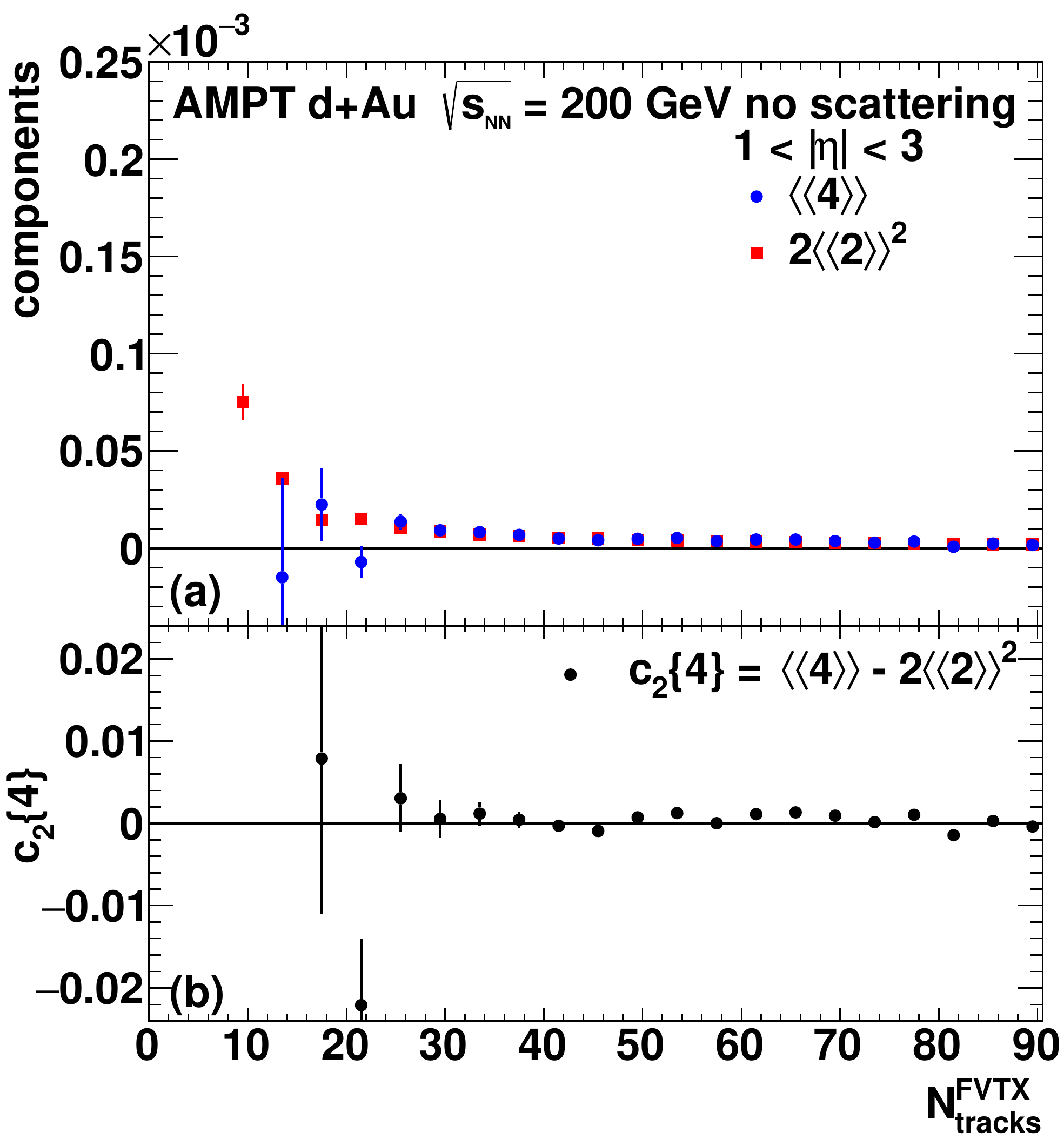}
\caption{Components $2\dmean{2}^2$ and $\dmean{4}$ (upper panels) and cumulant
  $c_2\{4\}$ (lower panels) as a function of $\nfvtxt$ in d+Au collisions at
  200 GeV.  The left plot is the data, the middle plot is AMPT, and the right
  plot is AMPT with scattering turned off.}
\label{fig:compandampt}
\end{figure}

To corroborate the interpretation of real valued $\vtf$ as indicative of
collective behavior, Figure~\ref{fig:compandampt} shows the individual components
$2\dmean{2}^2$ and $\dmean{4}$ (upper panels)
and the cumulant $\ctf = \dmean{4} - 2\dmean{2}^2$ (lower panels)
in the data (left plot), AMPT (middle plot),
and AMPT with no scattering (right plot).  When turning off the scattering,
the transport from initial geometry to final state interactions doesn't take
place, so that only non-flow effects remain.  Comparing the middle plot and
the right plot, the case with scattering shows a negative $\ctf$,
meaning real-valued $\vtf$, whereas the case without scattering shows
a $\ctf$ that's near zero but positive, indicating complex-valued $\vtf$.

\section{Summary}
\label{sec:summary}

In summary, we have presented measurements of multiparticle correlations
in d+Au collisions at 200, 62.4, 39, and 19.6 GeV.  We find real-valued
$\vtf$ for all energies, providing strong evidence of
collective behavior in small systems at RHIC energies down to the lowest energies.





\bibliographystyle{elsarticle-num}
\bibliography{refs}







\end{document}